\documentclass[journal]{IEEEtran}
\usepackage{amsmath,amssymb}
\usepackage{graphicx}
\usepackage{cite}
\usepackage{amsfonts}
\usepackage{booktabs}
\usepackage{multirow}
\usepackage{flushend}
\usepackage{mathrsfs}
\usepackage{hyperref}
\usepackage{cleveref}

\usepackage{longtable, booktabs}
\usepackage{supertabular}
\usepackage{lscape}
\usepackage{eurosym}
\usepackage{verbatim}
\usepackage{soul, color, xcolor}
\ifCLASSINFOpdf
\else
\fi

\begin{document}

\title{Sustainable Wireless Networks via\\ Reconfigurable Intelligent Surfaces (RISs):\\ Overview of the ETSI ISG RIS}

\author{
Ruiqi Liu, Shuang Zheng, Qingqing Wu, Yifan Jiang,  Nan Zhang, Yuanwei Liu, \\Marco Di Renzo, and George C. Alexandropoulos

\thanks{
R. Liu, S. Zheng, and N. Zhang are with the State Key Laboratory of Mobile Network and Mobile Multimedia Technology, Shenzhen, China and Wireless and Computing Research Institute, ZTE Corporation, Shenzhen, China (e-mail: \{richie.leo, zheng.shuang, zhang.nan152\}@zte.com.cn).}
\thanks{Y. Jiang and Q. Wu are with the Department of Electronic Engineering, Shanghai Jiao Tong University, Shanghai 200240, China (e-mail: yc27495@umac.mo; qingqingwu@sjtu.edu.cn).}
\thanks{Y. Liu is with the School of Electronic Engineering and Computer Science, Queen Mary University of London, UK (e-mail: yuanwei.liu@qmul.ac.uk).}
\thanks{M. Di Renzo is with Universit\'e Paris-Saclay, CNRS, CentraleSup\'elec, Laboratoire des Signaux et Syst\`emes, 3 Rue Joliot-Curie, 91192 Gif-sur-Yvette, France (e-mail: marco.direnzo@l2s.centralesupelec.fr).}
\thanks{G. C. Alexandropoulos is with the Department of Informatics and Telecommunications, National and Kapodistrian University of Athens, Panepistimiopolis Ilissia, 15784 Athens, Greece (e-mail: alexandg@di.uoa.gr).}
%\thanks{Corresponding author: G. C. Alexandropoulos.}
}

% make the title area
\maketitle
\begin{abstract}
Reconfigurable Intelligent Surfaces (RISs) are a novel form of ultra-low power devices that are capable to increase the communication data rates as well as the cell coverage in a cost- and energy-efficient way. This is attributed to their programmable operation that enables them to dynamically manipulate the wireless propagation environment, a feature that has lately inspired numerous research investigations and applications. To pave the way to the formal standardization of RISs, the European Telecommunications Standards Institute (ETSI) launched the Industry Specification Group (ISG) on the RIS technology in September 2021. This article provides a comprehensive overview of the status of the work conducted by the ETSI ISG RIS, covering typical deployment scenarios of reconfigurable metasurfaces, use cases and operating applications, requirements, emerging hardware architectures and operating modes, as well as the latest insights regarding future directions of RISs and the resulting smart wireless environments.
\end{abstract}


\IEEEpeerreviewmaketitle

\section{Introduction}
As wireless systems evolve through generations, more advanced capabilities are expected to support the ever-growing needs of end users in various communication applications, including vertical industries, which are usually accompanied with a cost of higher energy consumption and deployment costs. Towards the 6th Generation (6G), candidate technologies that are able to boost the performance of wireless connectivity in both cost-effective and energy-efficient ways are lately being developed, with one of the prominent ones being the Reconfigurable Intelligent Surfaces (RISs) \cite{WCM_RIS_Standards}. This technology deals with planar surfaces comprising large numbers of metamaterials with tunable responses over their impinging signals, being thus capable to shape the their reflection amplitude and phase shift. By intelligently coordinating the scattering of all metamaterials, an RIS is capable to actively reconfigure wireless channels contributing to the  energy efficiency of communications while increasing the coverage areas.

To conduct gap analysis for the RIS technology towards formal standardization, investigate the resulting open problems, and contribute potential solutions, a six-year roadmap has been set by the European Telecommunications Standards Institute
(ETSI) launching the Industry Specification Group (ISG) on RISs to investigate the technology's establishment, with the goal to  publish deliverables in the form of Group Reports (GRs). During this group's first phase from 2022 to 2023, analysis of the technology potential, validation, maturity timelines, and requirements for standardization were conducted. Initial guidelines on the RIS technology with possible group specifications (GSs) on its functional architecture are planned within 2024 and 2025. The third and final phase of the ISG, spanning 2026 to 2027, will mark the maturity of the various RIS solutions with the ISG planning to go deeper into standardization on practical matters of the technology, such as the ElectroMagnetic (EM) compatibility which is critical for co-existence and interference management.

In this article, the conducted and ongoing work of the ETSI ISG RIS is overviewed in conjunction with the envisaged activities for the near future. The currently identified key deployment scenarios and use cases are discussed in the following Section~\ref{Sec:deployment}. The available models for RIS functionality and the resulting RIS-parametrized channels are included and analyzed in Section~\ref{Sec:RIS_model}. Potential reference architectures with RISs are highlighted in Section~\ref{Sec:architecture} along with discussions on their potential impact in upcoming standards. The group's ongoing work as well as planned future directions are presented in Sections~\ref{Sec:ongoing_WIs} and~\ref{Sec:future_work}, respectively. Finally, Section~\ref{Sec:conclusions} includes the article's concluding remarks. 

\section{Deployment Scenarios and Use Cases}\label{Sec:deployment}
Before diving into the technical details of the RIS technology, the ISG RIS focused on identifying where RISs are needed and how they can be deployed to boost the performance of current and future communication systems. These directions are being investigated in GR $001$ \cite{GR_001}. In particular, this report emphasizes on three typical deployment scenarios: indoors, outdoors, and mixed indoors and outdoors.
\begin{figure*}[!t]
    \centering
    \includegraphics[width=0.96\textwidth]{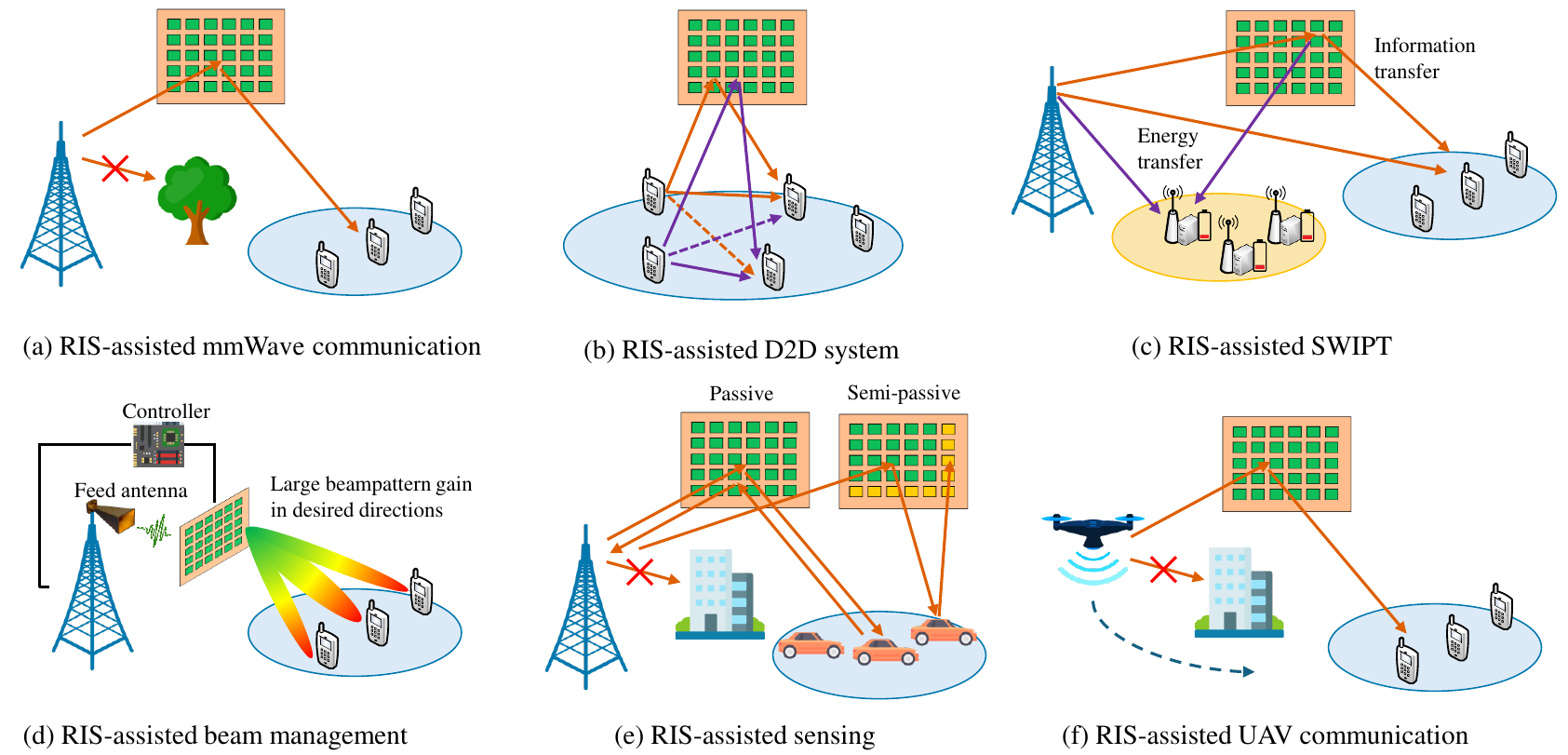}
    \caption{Illustration of the promising key use cases of RISs as identified in GR $001$~\cite{GR_001}.}
    \label{fig::use_cases}
\end{figure*}

For purely indoor deployments, the Line-Of-Sight (LOS) path may not exist in some areas, where RISs can be deployed instead of a more expensive and power hungry micro Base Station (BS). Indoor scenarios, such as automated industrial factories, may also benefit from the improvements in localization- and sensing-related capabilities an RIS can offer~\cite{ABoI_EURASIP}. The main challenge of network optimization in an outdoor environment is to provide seamless coverage in the entire cell under the existence of obstacles, such as buildings. RISs can be deployed at the top of rooftops or on the facade of buildings to help extend coverage in dense urban areas. In addition, User Equipment (UEs) in outdoor settings may have more significant mobility, and thus, dynamic RISs are needed for fast beam tracking. The third typical scenario is mixed indoors and outdoors, where RISs can play several roles. Indoor areas are often seen as weak coverage areas since BSs are usually deployed outdoors, and the conventional solution involves adding micro BSs or relays indoors to provide signal coverage. With RISs, it is possible to directly reflect or refract the signal from outdoors to serve indoor users. The transparent RIS, which can be installed as a layer of a window, stands out in this scenario as a promising solution. Some trial results in the three aforementioned scenarios, as reported by ISG RIS members, are described in \cite{9955484}.

%\section{Use Cases}\label{Sec:use_cases}
The various use cases capitalizing on the potential RIS deployment scenarios drive the definition of RIS physical models and channel models. So far, eight major types of use cases have been defined in \cite{GR_001}, as examples of how RISs can benefit network operators, end devices and users, as well as vertical industries. 
Typically, when obstacles exist in a wireless environment, especially for higher frequencies such as in the millimeter Wave (mmWave) band, the communication channel between the Transmitter (TX) and the Receiver (RX) can be blocked or shadowed. This is a case where RISs can come into play to overcome the problem. In Fig.~\ref{fig::use_cases}(a) and (b) examples use cases are included to illustrate how RISs can provide an artificial LOS path between communication nodes when natural ones do not exist. 

As the standardization for ambient Internet-of-Things (IoT) devices progresses, future wireless networks may serve simultaneously as information distributors and power sources. In the context of simultaneous wireless information and power transfer, as illustrated in Fig. \ref{fig::use_cases}(c), RISs can contribute to achieve finer real-time power allocation as well as larger diversity gain. Since 6G is expected to require significantly larger bandwidths for both communications and sensing, it is clear that higher frequency bands will be needed, which imposes challenges to conventional massive Multiple-Input Multiple-Output (MIMO) architectures. Continuing using this architecture means a drastic increase in the manufacturing cost and power consumption as the number of antenna elements will increase. However, if RISs are deployed close to the BS to assist on achieving beam management, a simple feed antenna can be used at the BS side.
What's more, RISs can also empower and strengthen novel network functions in 5G-advanced and 6G, including sensing passive objects. 

\section{Modeling of RISs and RIS-Aided Channels}\label{Sec:RIS_model}
The GR RIS $003$ that was published in June $2023$ mainly includes~\cite{GR_003}: \textit{i}) models for the behavior of the unit elements of reconfigurable metasurfaces as well as their overall structure, which aim at a suitable trade-off between EM accuracy and simplicity for performance evaluation and optimization of the RIS technology at different frequency bands; and \textit{ii}) end-to-end channel models, parametrized over the RISs incorporated within the wireless environment, which target at accurately describing the impact of metasurfaces on signal propagation. In addition, the report discusses: \textit{iii}) channel estimation approaches for wireless systems incorporating solely reflecting and almost passive RISs as well as semi-passive RISs equipped with sensing units~\cite{alexandropoulos2023hybrid} (either performing sensing simultaneously with tunable reflection or in an orthogonal manner via separate time slots); and \textit{iv}) key performance indicators and methodologies for evaluating the performance of RISs for applications of wireless communications.

The locally periodic discrete model of an RIS considers periodic boundary conditions at the RIS unit-cell level, which is accordingly associated with a set of complex-valued coefficients. Each of those coefficients, which indicates the ratio between the reflected and incident electric fields for an infinite RIS with the same respective unit-cell configuration (i.e., the reflection coefficient), is obtained by appropriately configuring the electronic circuits of the unit cell. Following this line of modeling, the RIS structure is modeled as a homogeneous surface that realizes specular reflection. The amplitude of each reflection coefficient is bounded by unity for an RIS without reflection amplification units~\cite{RISsurvey2023}, however, it may depend on its respective phase, and thus, they need to be jointly optimized. 
\begin{figure}[t]
    \centering
    \includegraphics[width=\columnwidth]{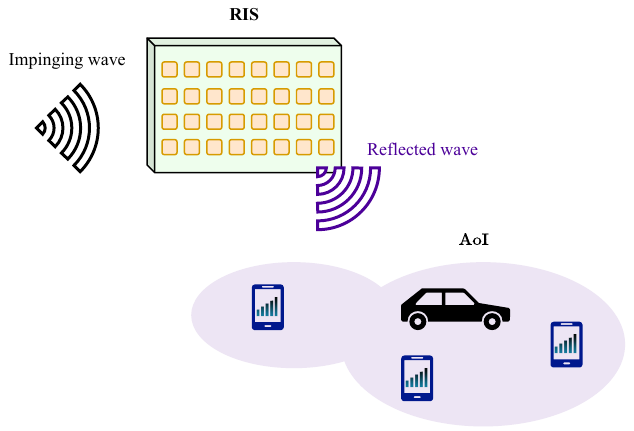}
    \caption{The Area of Influence (AoI) of an RIS is defined as the geographical area for which a quality-of-service threshold is satisfied with the aid of an RIS, i.e., \( {\rm AoI} \triangleq \{ \mathcal{S}|m(\mathcal{S})\geq q_{\rm th} \}\) with \(m(\cdot)\) denoting the desired performance metric, \(\mathcal{S}\) is the geographical space, and \(q_{\rm th}\) is the performance threshold~\cite{ABoI_EURASIP}. The AoI depends on the physical characteristics of the RIS, and given those characteristics, it can be dynamically adjusted.}
    \label{fig:AoI}
\end{figure}

An RIS whose unit-cells' sizes and their inter-distances are much smaller than the wavelength can thus be modeled by its effective surface parameters, which appear in EM problems that are formulated as effective boundary conditions (polarizabilities, susceptibilities, or impedances/admittances). Under these assumptions, the RIS can be modeled as an inhomogeneous sheet of polarizable particles (i.e., unit cells) that is characterized by an electric surface impedance and a magnetic surface admittance, which, for general wave transformations, are dyadic tensors. These two dyadic tensors constitute the macroscopic homogenized model of an RIS. Once the homogenized and continuous electric surface impedance and magnetic surface admittance are obtained based on the desired wave transformations, the microscopic structure and physical implementation of the RIS in terms of unit cells are obtained. Once the macroscopic surface impedance and admittance are determined, appropriate geometric arrangements of sub-wavelength unit cells and the associated tuning circuits that exhibit the corresponding electric and magnetic response are characterized by, typically, using full-wave EM simulations.
\begin{figure}[t]
    \centering
    \includegraphics[width=\columnwidth]{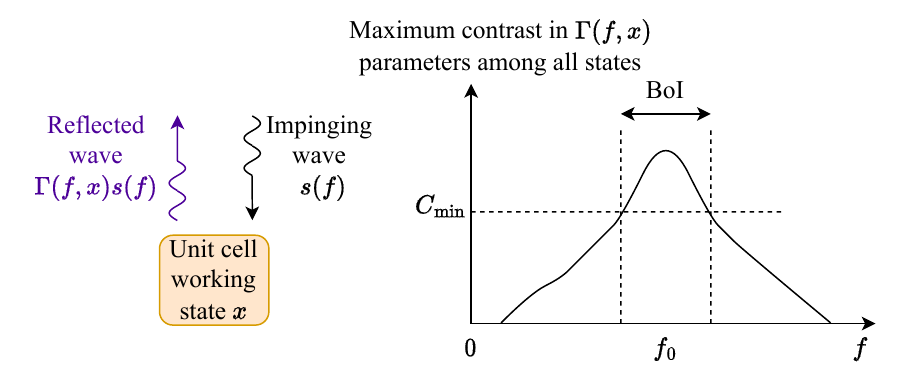}
    \caption{Let \(\Gamma(f,x)\) denote the reflection coefficient of an RIS unit cell, given its working configuration \(x\) and the operating frequency $f$, applied to an impinging wave \(s(f)\). To characterize the Bandwidth of Influence (BoI) of an RIS consisting of multiple identical unit cells around a targeted central frequency $f_0$, the maximum contrast among all different states \(x\) of a single unit cell, defined as: \(C(f)\triangleq \max_{\forall x,x\neq x^{'}} |\Gamma(f,x)-\Gamma(f,x^{'}) |\), has been lately used~\cite{ABoI_EURASIP}, as shown the right graph. The BoI is defined as the set of frequencies where the maximum contrast is above a threshold value $C_{\rm min}$. If the maximum difference between any two different states tends to zero, the RIS offers no reconfigurability and acts as a dummy reflector with fixed frequency characteristics depending on the material used and the angle of incidence.}		
    \label{fig:BoI}
\end{figure}

According to GR RIS $003$, the wireless channels between a TX and an RIS as well as between an RIS and an RX can be described by typical 3rd Generation Partnership Project (3GPP) channel models. However, special care has been given for modeling the pathloss in RIS-aided communications, which can quite frequently take place in the near-field regime. This happens for large RIS sizes and high frequencies, i.e., millimeter waves and above. A near-field pathloss model, expressed as a function of the transmission power, the
gain of the RIS unit cells, the square of the wavelength, and the received power, was introduced, indicating that the received power is related to the normalized power radiation patterns of the TX/RX antennas and RIS elements, their reflection coefficients, and the distances between the TX/RX and the RIS
elements. In addition, a far-field pathloss model, for the case where the signals reflected by all the RIS elements are aligned in phase to enhance the received signal
power, was presented. An empirical pathloss model for a TX-RIS-RX channel was also described as well as a fully stochastic (i.e., unstructured) and a stochastic geometrical (i.e., structured) models for sub-6 GHz and a structured model for high frequencies. A model for RISs with dual polarization has been also presented.

Depending on how an RIS is realized, it can re-radiate multiple waves for a given incident electromagnetic wave. Those unwanted scattered waves should be considered as interference, i.e., undesired waves for the targeted wireless communications. A model that captures this behaviour has been presented in the report. Other forms of interference in RIS-empowered communication systems are: \textit{i}) interference generated by other RIS devices within an operator's network; and \textit{ii}) interference generated by RIS devices of other operators' networks. In addition, an RIS can be deployed in a malicious way to enable eavesdropping~\cite{PLS2022_counteracting_all}, thus, enabling, possibly transparent to the legitimate nodes, unwanted links. To this end, the report has adopted and described in detail the notions of the RIS Area of Influence (AoI) (see Fig.~\ref{fig:AoI}) and Bandwidth of Influence (BoI)~\cite{ABoI_EURASIP} (sketched in Fig.~\ref{fig:BoI}), which intend to quantify the RIS impact in the spatial and frequency domains, respectively, while constituting performance indicators that can be realized in synergy by RIS hardware and algorithmic designs.

\section{Reference Architecture and Standardization}\label{Sec:architecture}
Some reference architectures for the deployment of RISs in wireless networks are presented in GR RIS $002$~\cite{GR_002}. One possibility is to integrate RISs to replace phase shifters and power amplifiers in conventional massive MIMO transmitters with the goal to reduce cost and power consumption. With this deployment option, the RIS is essentially a part of the TX unit. The predominant consideration, however, is to deploy RISs as relays in distributed locations within the cell, with the goal to enhance signal strength in weak coverage areas. 

\begin{figure}[t]
\centering
\includegraphics[width=0.9\linewidth]{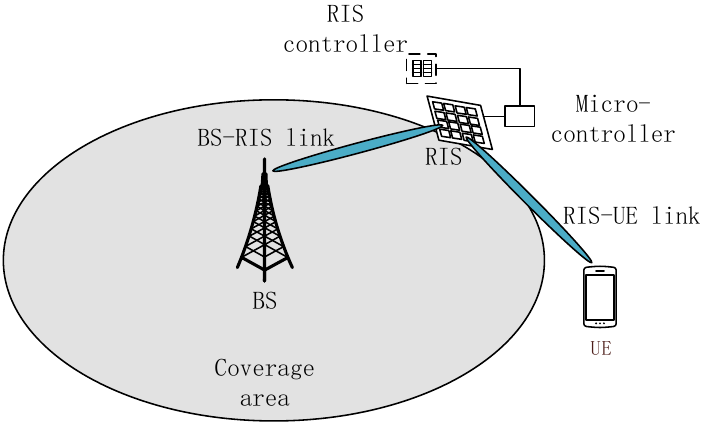}
\caption{A reference deployment architecture for RIS-assisted cellular networks. The RIS extends the coverage area of the BS when optimized to focus wireless propagation in the intended UE position.}
\label{fig:arch}
\end{figure}

In the reference deployment architecture in Fig. \ref{fig:arch}, for the downlink case, the RIS reflects the signal from BS to UE, creating effectively two channels: the BS-RIS and RIS-UE links. For the uplink case, signal propagation take place vice versa. A micro-controller is connected to the RIS panel to configure in real-time the responses of the metasurface's reflective elements, the so-called RIS reflection coefficients, and it receives control signals from the RIS controller~\cite{ABoI_EURASIP}. The latter enables the interactions of the RIS with the BS or the UE, according to the deployment scenario and use case, receiving control signals from any of them. %Depending on which device to serve as the RIS controller, GR RIS 002 \cite{GR_002} further identifies the network-controlled mode, 
In the network-controlled mode defined in GR RIS $002$~\cite{GR_002}, the network (comprising the BS and the core network) determines the control information. Since the network is able to collect and utilize system information of potentially multiple network nodes including BSs, UEs, and RISs, this mode enables the realization of a potential globally optimized resource allocation scheme and overall beamforming configurations, facilitating interference mitigation. On the other hand, it is also possible that the RIS controller receives control information from the UEs, under cases such as the sidelink communication between UEs. To achieve a global optimum RIS-enabled network configuration, the network may transmit assisting information to the UE to facilitate the decision on how to control the RIS. Either way, to fully reap the benefits of RISs and ensure a harmonized smart wireless environment, it is expected for each RIS controller to have signaling exchange with other network nodes. 

\begin{figure*}[!t]
\centering
\includegraphics[width=0.9\textwidth]{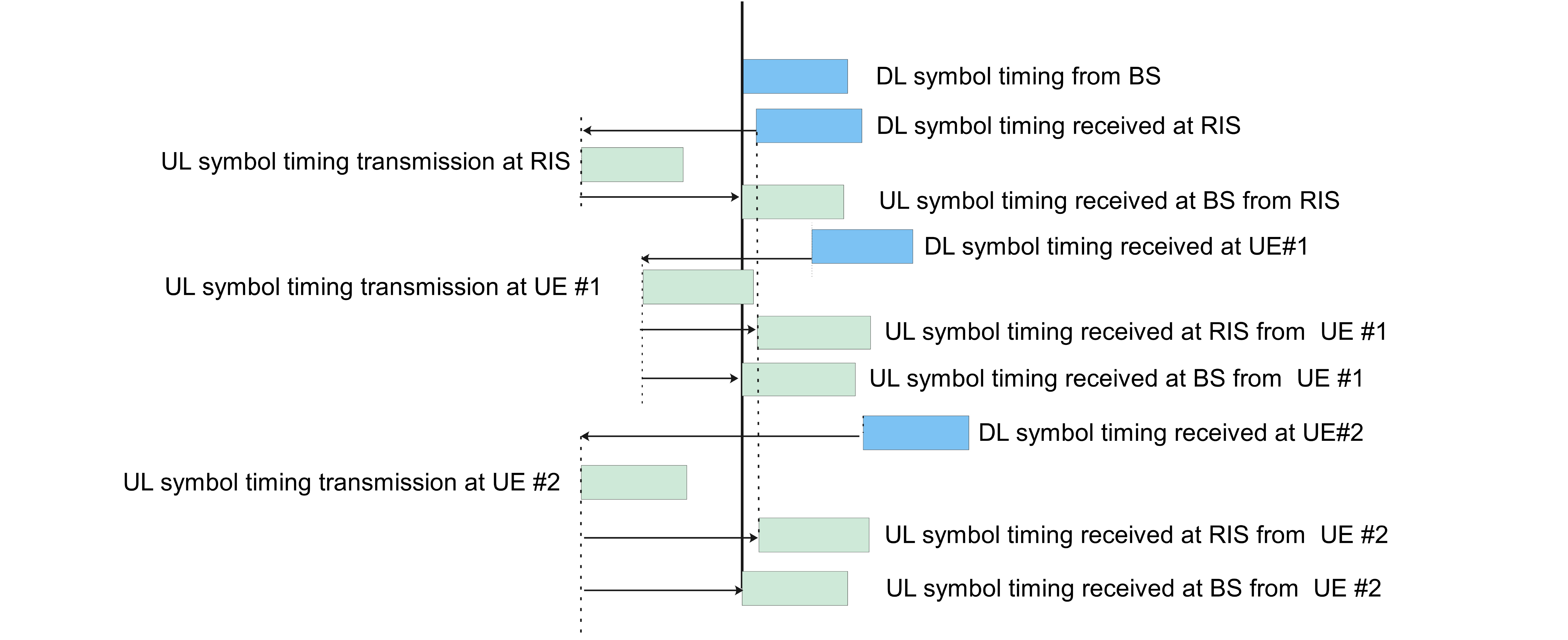}
\caption{A timing advance example in a wireless network comprising one BS and two UEs, with all profiting from the deployment of a single RIS. The timing information can be used from the system to distinguish the RIS utilization between the two UEs.}
\label{fig:timing}
\end{figure*}

As previously discussed, with the integration of RISs into the wireless communication network, control information needs to be exchanged via the air interface between any RIS controller and the network or between any RIS controller and the UE(s). It is anticipated that a significant impact on standards will take place for RIS-integrated wireless systems. One possible impact is that multiple diverse information, including spatial information, on/off state information, timing information, operating mode information, etc., may need to be exchanged via the BS-RIS or the RIS-UE link for controlling the RIS operation. The spatial information can be the RIS beam (or phase profile/configuration) information that is indicated based on the existing Transmission Configuration Indicator (TCI) state or can be a set of phases and amplitudes for all reflective elements to configure the RIS. The timing information, with an example shown in Fig. \ref{fig:timing}, can be leveraged when the RIS is deployed away from the BS, as follows. With the timing advance adjustment information from the BS, the uplink symbols transmitted from the UE\#1 and UE\#2 can be received at the RIS and BS side at the same time. Besides, considering that multiple RISs may be integrated in future smart wireless communication systems (similar to distributed MIMO systems)~\cite{ABoI_EURASIP}, the on/off information can be indicated explicitly or implicitly to turn on/off the RISs for purposes related to energy saving and interference elimination. 

The exchange of the latter RIS-related control information is expected to have an impact to standards.  For the network-controlled RIS, the control information can be transmitted from the BS to the RIS via the existing Uu interface (e.g., the Physical Downlink Control Channel (PDCCH) or the Physical Downlink Shared Channel (PDSCH)), while for the UE-controlled RIS, the control information can be transmitted from the UE to the RIS via the Uu interface, the PC5 interface (e.g., the Physical Sidelink Control Channel (PSCCH) or the Physical Sidelink Shared Channel (PSSCH)), or a private interface. When dealing with an RIS capable of UL transmissions, the feedback information (e.g., the acknowledgment or negative acknowledgment (ACK-NACK) for the received control information) can be transmitted from the RIS to the BS via the Uu interface(e.g., the UL Physical Uplink Control Channel (PUCCH)) or transmitted from the RIS to the UE via the PC5 interface. 

Apart from the latter options for realizing RIS-related control information, enhancements in the current signaling structure may be needed. It is thus critical for the standards' development organizations, such as 3GPP, to start investigating how to incorporate the RIS novel aspects as well as the relevant potential updates to the current standards.

\section{Ongoing Work} \label{Sec:ongoing_WIs}
Besides the aforementioned three published GRs from the ETSI ISG RIS, there are also four new ongoing Working Items (WIs) studying the implementation, diversity and multiplexing, plurality of functionalities, and near-field channel modeling of RISs, which are summarized in Table~\ref{ongoing_WIs}. In the following, we discuss the latest progress, at the time of writing this article, within each of these items.

Having started in February 2023, the DGR-$004$~\cite{DGR_004} focuses on the practical implementation and considerations of RISs. Up to now, this item has nearly reached a `stable draft' status. It introduces the general hardware aspects of RISs, including the unit-cell operation principle, metasurface design, reconfigurable elements, RIS controller, and the RIS power source. In addition, design requirements and practical implications for reflective, refractive, and absorptive RISs along with relevant extensive simulation results are discussed. Some initial trial and measurement results using available RIS prototypes are provided, which are also evaluated with respect to the RIS cost, complexity, and energy-efficiency. All in all, DGR-004 provides a comprehensive and useful guidance on the hardware design and implementation of RISs.

The DGR-005~\cite{DGR_005} was launched in May 2023 aiming to provide a comprehensive analysis on the diversity and multiplexing gain that can be achieved with RIS-aided wireless communications. For the diversity schemes with RISs, this item is planned to share the perspectives from the time, frequency, transceivers, polarization, and spatial domains. For the multiplexing schemes associated with RISs, it will discuss the spatial-division and other possible multiplexing modes. For each category, the impact of channel estimation on the diversity and multiplexing potential as well as corresponding performance gain comparisons will be conducted. Moreover, the RIS performance gains in synergy with other multi-antenna technologies, such as beamforming and antenna array selection, will also be identified.

In parallel to DGR-005~\cite{DGR_005}, in May 2023, the DGR-006~\cite{DGR_006} was also initiated. This item targets to explore RIS advances in the direction of metasurfaces with dual or other than reflection functionalities, termed as Multi-Functional (MF)-RISs. Simultaneously transmitting and reflecting (STAR)-RISs and RISs with sensing capabilities are the two types of MF-RISs that are expected to be investigated in detail~\cite{RISsurvey2023}. For each of these categories, and any others that may appear, DGR-006 is responsible for introducing basic signal and channel models, operating protocols, reconfigurable coefficient designs, deployment strategies, resource allocation schemes, and performance analysis. Interestingly, the MF-RISs that will be presented in DGR-006 are expected to impact other ISGs, such as the ISG on Integrated Sensing and Communications (ISAC) and on TeraHertz (THz) modeling.

Very recently, in March 2024, the DGR-007 \cite{DGR_007} started its activity on the investigation of near-field propagation caused by RISs with large aperture sizes and high operating frequencies, complementing the relevant existing study in DGR-003. This item is tasked to provide specific use cases and scenarios, channel models, tailored technical solutions, and impact analysis of RIS-aided near-field wireless systems.
\begin{table*}
	\begin{center}
		\caption{Summary of the new ongoing WIs in ETSI ISG RIS.}
		\label{ongoing_WIs}
		\begin{tabular}{|p{50pt}|p{65pt}|p{240pt}|p{60pt}|}
			\hline
			Identifier & Title & Main objectives & Estimated publication time \\
			\hline
			DGR-$004$~\cite{DGR_004} &  Implementation and Practical Considerations & Investigation of implementation and practical considerations for RISs in a wide range of frequency bands and deployment scenarios; consult possible solutions and prototyping results. &  August 2024 \\
			\hline
			DGR-$005$~\cite{DGR_005} & Diversity and Multiplexing of RIS-aided Communications & Identification of use cases and deployment scenarios of RIS-based diversity and multiplexing schemes. Investigation of feasible diversity and multiplexing schemes for different RIS hardware and operating modes, while identifying the TX/RX complexity increase or reduction for the various multiplexing schemes. Evaluate the characteristics of RIS-aided channels for different diversity schemes and analyze the performance gains and impact. & July 2024   \\
			\hline
		   DGR-$006$~\cite{DGR_006} & Multi-functional Reconfigurable Intelligent Surfaces (RIS): Modelling, Optimisation, and Operation & Identification of challenges and inclusion of efficient solutions for Multi-Functional RISs (MF-RISs). Cchannel modelling, phase profile optimization, resource allocation, and practical deployment schemes for MF-RISs.  &September 2024 \\
     \hline
		   DGR-$007$~\cite{DGR_007} & Near-field Channel Modeling and Mechanics & Identification of use cases and scenarios for near-field RIS-aided wireless systems and related channel modeling, including methods to extend to the near-field region the available cascaded channel modeling. Investigation of technical solutions/mechanisms applicable for RIS-asssted wireless systems in the near field; identification and analysis of potential specification impact to stardards.  & December 2025  \\
			\hline
		\end{tabular}
	\end{center}
\end{table*}

\section{Future Directions on the RIS Technology} \label{Sec:future_work}
RISs have been already recognized by many standardization bodies, including the International Telecommunication Union (ITU), as a promising candidate wireless technology. In a report published by the ITU on future technology trends for 6G \cite{liu_2023_6G}, RIS is described as an effective method to support evolved MIMO, to improve the performance and overcome the challenges in traditional beam-space antenna array beamforming, and to enable higher frequency ranges such as sub-THz. It is worth noting that RISs also take up a dedicated $2$-page subsection in ITU's $44$-page report on wireless networks covering aspects from emerging applications to technology enablers in radio interfaces as well as network architectures.

In parallel, it has been very recently reported that RISs can be integrated into the 5G and 5G-advanced networks to boost network performance in a variety of deployment scenarios~\cite{9955484,9999288}. The results therein obtained by field trials demonstrate the feasibility to adopt RISs in cellular networks, even under sophisticated environments.

To meet the expectation for RIS adoption in 6G, it is vital for 3GPP to formally specify RISs as a new network component. This will ensure the interoperability between RISs and other network nodes. In 3GPP Release $18$, the Network-Controlled Repeater (NCR) has been established as a WI, which aims to enhance over conventional repeaters to support the capability or receiving and processing side control information from the network. NCRs are also designed to support beamforming when transmitting and receiving, with the goal to boost spectral efficiency. 

Considering that part of the functionalities of RISs are similar to NCRs, it is important to investigate the NCR functionalities that can be reused by RISs without additional effort, and which functionalities are specific to RISs. Those will naturally require additional studies towards standardization. In addition, the 3GPP Release $19$ has recently launched channel modeling work for ISAC and for new spectrum. This activity can be leveraged for RISs thanks to some similarities in the modeling, such as the reflection and multi-path propagation. Therefore, identifying the similarities and differences between RISs and existing normative work is of great significance to promoting the standardization work for the RIS technology. 

\section{Conclusion}  \label{Sec:conclusions}
The ISG RIS is the first ETSI group dedicated to the investigation, evolution, and standardization of the RIS technology, which undoubtably constitutes a promising candidate technology for 6G networks. In this article, we overviewed the work carried out up to date in the ETSI ISG RIS, including deployment scenarios, use cases, physical and channel models, as well as potential architecture and impact to standards. We have also discussed ongoing work in the group dealing with practical considerations of the technology, diversity and multiplexing schemes, MF-RISs, as well as near-field models and mechanics.  
Interested readers are encouraged to read further the ISG's GRs, which are provided as the references of this article, and to participate to the group's fascinating ongoing and future activities.

\section*{Acknowledgement}
This work has been partially supported by the SNS JU project TERRAMETA under the EU's Horizon Europe research and innovation programme under grant agreement number $101097101$, including top-up funding by UKRI under the UK government's Horizon Europe funding guarantee.

\bibliographystyle{IEEEtran}
\bibliography{Reference}

\end{document}